\documentclass[10pt]{iopart}
\usepackage{graphicx}
\usepackage{iopams}

\newcommand{\Erfc}{\mathop{\mathrm{Erfc}}\nolimits}
\newcommand{\sgn}{\mathop{\mathrm{sgn}}\nolimits}

\newcommand{\abs}[1]{|#1|}
\newcommand{\binom}[2]{\left(\begin{array}{c}#1\\#2\end{array}\right)}
\newcommand{\vect}[1]{\bi{#1}}
\newcommand{\journaltitle}[1]{\textit{#1}}
\newcommand{\booktitle}[1]{\textit{#1}}
\newcommand{\volume}[1]{\textbf{#1}}

\begin{document}
\jl{1}

\title[Path-decomposition expansion and edge effects]{Path-decomposition 
expansion and edge effects in a confined magnetized free-electron gas}
\author{M M Kettenis and L G Suttorp}
\address{Instituut voor Theoretische Fysica, Universiteit van Amsterdam, 
Valckenierstraat 65, 1018 XE Amsterdam, The Netherlands}

\begin{abstract}
Path-integral methods can be used to derive a `path-decomposition expansion' 
for the temperature Green function of a magnetized free-electron gas confined 
by a hard wall. With the help of this expansion the asymptotic behaviour of 
the profiles for the excess particle density and the electric current density 
far from the edge is determined for arbitrary values of the magnetic field 
strength. The asymptotics are found to depend sensitively on the degree of 
degeneracy. For a non-degenerate electron gas the asymptotic profiles are 
essentially Gaussian (albeit modulated by a Bessel function), on a length 
scale that is a function of the magnetic field strength and the temperature. 
For a completely degenerate electron gas the asymptotic behaviour is again 
proportional to a Gaussian, with a scale that is the magnetic length in this 
case. The prefactors are polynomial and logarithmic functions of the distance 
from the wall, that depend on the number of filled Landau levels $n$. As a 
consequence, the Gaussian asymptotic decay sets in at distances that are large 
compared to the magnetic length multiplied by $\sqrt{n}$.
\end{abstract}   




\maketitle

\section{Introduction}

In a magnetized charged-particle system edge effects are of paramount 
importance. An illustration of this fact is furnished by the phenomenon of 
Landau diamagnetism \cite{LAN:1930}, which is due to electric currents flowing 
near the boundaries of the sample. As a further example one may mention 
quantum Hall systems for which the relevance of edge effects has amply been  
shown.

The analysis of the influence of the boundary on the properties of a quantum 
many-body system is a difficult mathematical problem. Even if the bulk 
properties of the unconfined system are understood, the presence of the edge 
leads to a boundary-value problem that is often difficult to solve 
analytically. Leaving out the interparticle interaction simplifies this 
problem quite a lot, although even in that case the analysis remains 
complicated. A physical system that is particularly relevant in this context 
is the non-interacting electron gas in a uniform magnetic field in the 
presence of a confining hard wall. In fact, this is the system in which Landau 
diamagnetism, with currents flowing near the edge, can be studied in its 
purest form.  

Several methods have been devised to analyse edge effects in the confined 
magnetized free-electron gas. At zero temperature one may try to solve the 
eigenvalue problem in terms of distorted Landau levels and determine the edge 
currents by summing the contributions of the lowest-lying eigenfunctions. Even 
for a simple flat geometry this leads to a rather involved mathematical 
analysis in terms of parabolic cylinder functions \cite{MAC:1984}, 
\cite{KUN:1994}. Recently, we studied the profiles of the particle density and 
the electric current density along these lines \cite{KES:1998}.

An alternative approach starts by focusing on the high-temperature regime, 
where Maxwell-Boltzmann statistics applies. In that case a convenient tool is 
furnished by the one-particle temperature-dependent Green function. As shown 
by Balian and Bloch \cite{BAB:1970} the Green function for the confined system 
can be related to that of the corresponding system without boundaries by 
making a systematic expansion that accounts for an increasing number of 
reflections of the particles against the confining wall. The ensuing 
multiple-reflection expansion was used in recent years \cite{ROB:1986}, 
\cite{JOS:1995} to determine the profiles of the particle density and the 
(electric) current density for small values of the magnetic field. These 
small-field profiles had been found before from perturbation theory 
\cite{OHM:1973}, \cite{JAN:1980}. It turns out to be difficult to generalize 
these results for the profiles to arbitrary field strength and to relate them 
to those obtained by means of the eigenvalue method.

Some time ago Auerbach and Kivelson \cite{AUK:1985} invented a path-integral 
method to analyse boundary effects in Green functions. By suitably decomposing 
the relevant paths near the edge they derived a so-called `path-decomposition 
expansion' (PDX) for the one-particle Green function. The aim of the present 
paper is to see whether the use of PDX may shed light on the difficulties 
mentioned above and whether it may lead to new results on the profiles of 
physical quantities for arbitrary field strength, both for the 
high-temperature region and in the regime of high degeneracy.

The plan of the paper is as follows. We start by a review of the 
path-decomposition expansion and its derivation from the Feynman-Kac path 
integral. Particular attention will be given to the convergence of the PDX 
series. It will be shown that a suitable resummation can greatly enhance that 
convergence. The connection with the multiple-reflection expansion will be 
established. Subsequently, the extension of the method so as to include 
magnetic fields will be discussed by starting from the Feynman-Kac-It\^o 
representation.

For the specific case of a non-interacting charged-particle system in a 
uniform magnetic field, and confined by a hard wall parallel to the field, the 
general form of the terms in the PDX series can be established in detail. That 
result will be used to determine the first few terms of the asymptotic 
expansion for the profiles of the particle density and the current density. 
This asymptotic expansion is valid far from the edge and in the 
high-temperature regime. In contrast to earlier work we will not need to 
restrict ourselves to small field strengths, as we shall establish the full 
field dependence of the profiles. As it turns out, the precise knowledge of 
the asymptotic profiles for high temperatures and arbitrary fields is 
essential in determining how the profiles for the degenerate case depend on 
the filling of the Landau levels. 

\section{Path-decomposition expansion}

Consider a particle in an external potential $V(\vect{r})$, i.e.\ with the 
Hamiltonian
\begin{equation}
\label{eq:Hamiltonian}
  H = \frac{p^2}{2} + V(\vect{r})
\end{equation}
where we have chosen units in such a way that the particle mass drops out. The 
equilibrium quantum statistical properties of a set of particles moving in the 
potential $V$ is governed by the temperature Green function $G_\beta 
(\vect{r}', \vect{r})$, with $\beta$ the inverse temperature. Its 
path-integral representation is given by the Feynman-Kac formula 
\begin{equation}
\label{eq:Feynman-Kac}
  G_\beta (\vect{r}', \vect{r}) = 
  \langle \vect{r}' | \rme^{-\beta H} | \vect{r} \rangle =
  \int \rmd \mu_{\vect{r}, 0}^{\vect{r}', \beta}(\omega) \exp\left[
    -\int_0^\beta \rmd\tau\, V(\omega(\tau))\right]
\end{equation}
where $\omega(\tau)$ describes the path and $\rmd\mu_{\vect{r},0}^{\vect{r}', 
\beta}$ is the conditional Wiener measure \cite{ROEP:1996}. 

If a hard wall confines the particles to a region of space, the potential can 
be written as $V(\vect{r}) = V_0(\vect{r}) + V_w(\vect{r})$, where $V_w$ is a 
steep wall potential and $V_0$ is a smooth external potential. Exact 
evaluation of (\ref{eq:Feynman-Kac}) for the confined problem is in general 
not possible, even if the corresponding unconfined problem can be solved 
completely. In this section we will explore the use of the so-called 
`path-decomposition expansion' (PDX), first introduced by Auerbach and 
Kivelson \cite{AUK:1985}, to determine the Green function of the confined 
problem.

To simplify matters, consider the one-dimensional case, with a hard wall at 
$x=0$, i.e. $V_w(x) = \infty$ for $x < 0$ and $V_w(x) =0$ for $x > 0$. As a 
first step, we split the Green function into two parts
\begin{equation}
  \label{eq:split}
  G_\beta (x', x) = G^0_\beta (x', x) + G^c_\beta (x', x). 
\end{equation}
Here $G^0_\beta$ is the Green function for the problem without a wall. In 
order to calculate it, one needs to specify the potential $V_0(x)$ for $x<0$ 
as well.  We will take the latter to be the analytical continuation of the 
potential for $x>0$. We shall assume that the resulting $V_0(x)$ is such that 
$G^0_\beta$ can be evaluated in closed form. 

The second term of (\ref{eq:split}) is the difficult part. It is a correction 
that contains contributions from all paths crossing the boundary at least 
once, with an additional minus sign so as to compensate the corresponding 
contributions in $G^0_\beta$. In order to calculate $G^c_\beta$, one 
discretizes the path integral in the usual way by introducing $n$ evenly 
spaced grid points at $\tau_m = m \epsilon_n$, with $\epsilon_n = 
\beta/(n+1)$. Subsequently, one decomposes the paths at the boundary 
\cite{AUK:1985}. Here `decomposing' means that the paths are split into two at 
the point $\tau$, where they cross the boundary for the last time.  Choosing 
this `point of no return' between $\tau_m$ and $\tau_{m+1}$, one writes the 
path integral for $G^c_\beta$ as
\begin{eqnarray}
\label{eq:correction}
\fl  G^c_\beta (x', x) = - \lim_{n\rightarrow\infty} \sum_{m=1}^{n}
  \int_{-\infty}^0 \rmd x_m \int_0^\infty \rmd x_{m+1}\nonumber\\
\times  G_{\beta-(m+1)\epsilon_n} (x', x_{m+1})
G^0_{\epsilon_n} (x_{m+1}, x_m) G^0_{m\epsilon_n} (x_m, x) 
\end{eqnarray}
for $x$ and $x'$ both positive.

In the small interval between $\tau_m$ and $\tau_{m+1}$ the potential $V_0$ 
can be ignored, since the error will vanish in the continuum limit. Hence, we 
may use in that interval the `free' propagator $G^0_{f,\beta}(x', x) 
=(2\pi\beta)^{-1/2} \exp [-(x'-x)^2/2\beta]$, where we have put $\hbar=1$. The 
free propagator satisfies the identity
\begin{equation}
\label{eq:identity}
\fl  G^0_{f,\beta}(\abs{x'}, -\abs{x}) = \sgn(x') \lim_{x'' \downarrow 0}
  \int_0^\beta \rmd\tau\,  \frac{\partial}{\partial
    x''} G^0_{f,\beta-\tau} (x', x'')\, G^0_{f,\tau} (0, x).
\end{equation}
This identity, which is a generalization of that used in \cite{AUK:1985}, 
follows directly by differentiation of the relation
\begin{equation}
\fl  \int_0^1 \rmd\tau [\tau(1-\tau)]^{-1/2} \rme^{-a^2/\tau-b^2/(1-\tau)}
  = \pi \Erfc(a+b) \quad\quad (a>0,\, b>0)
\end{equation}
with respect to $b$. Using (\ref{eq:identity}) with $x<0$ and $x'>0$ in 
(\ref{eq:correction}), we get
\begin{eqnarray}
\fl  G^c_\beta (x', x) = -\lim_{n\rightarrow\infty} \lim_{x''\downarrow
  0} \sum_{m=1}^{n} \int_0^{\epsilon_n} \rmd\tau \int_{-\infty}^0 \rmd x_m 
\nonumber\\
\times \frac{\partial}{\partial x''} G_{\beta-\tau-m\epsilon_n} (x', x'')
G^0_\tau(0, x_m) G^0_{m\epsilon_n} (x_m, x).
\end{eqnarray}
The integral over $x_m$ can be extended to the interval $[-\infty, \infty]$, 
if a compensating factor $1/2$ is inserted. In fact, only small values of 
$x_m$ contribute anyway, at least in the continuum limit, owing to the 
presence of the second $G^0$ function. For these small values of $x_m$ the 
integrand is approximately invariant under a change of sign of $x_m$. 
Subsequently, we may join the two $G^0$ into one, so that we get a closed 
integral relation
\begin{equation}
  \label{eq:PDX_I}
  G^c_\beta (x', x) = - \lim_{x'' \downarrow 0} \frac{1}{2}
  \int_0^\beta \rmd\tau\,  \frac{\partial}{\partial
  x''}  G_{\beta-\tau} (x', x'')\, G^0_\tau (0, x)
\end{equation}
for positive $x$ and $x'$. This integral relation is the PDX formula derived 
in \cite{AUK:1985}. Since the right-hand side contains the original Green 
function $G$, we can iterate this integral equation by inserting 
(\ref{eq:split}). In this way we arrive at the PDX series:
\begin{eqnarray}
  \label{eq:MRE_I}
\fl  G_\beta (x', x) = G^0_\beta (x', x) 
- \lim_{x''\downarrow 0} \frac{1}{2} \int_0^\beta \rmd\tau\,
   \frac{\partial}{\partial x''}
  G^0_{\beta-\tau} (x', x'')\, G^0_\tau(0, x) \nonumber\\
+ \lim_{x''\downarrow 0} \left(\frac{1}{2}\right)^2 \int_0^\beta
  \rmd\tau\, 
  \frac{\partial}{\partial x''}
  \lim_{x'''\downarrow 0} \int_0^{\beta-\tau} \rmd\tau' \nonumber\\
\times  \frac{\partial}{\partial x'''}
   G^0_{\beta-\tau-\tau'} (x', x''')\, G^0_{\tau'} (0, x'')
   \, G^0_\tau (0, x) - \ldots
\end{eqnarray}

To study the convergence of the PDX series we look at the special case of a 
vanishing external potential $V_0$.  The Green function for a free particle in 
the presence of a hard wall can be calculated using a reflection principle. It 
reads
\begin{equation}
  \label{eq:G_free}
  G_{f,\beta} (x', x) = G^0_{f,\beta} (x', x) - G^0_{f,\beta} (-x', x)
\end{equation}
for $x$ and $x'$ positive. Since the first term is $G^0_\beta$, the second 
term must be the correction $G^c_\beta$. To check the validity of 
(\ref{eq:MRE_I}) for the present case, we employ (\ref{eq:identity}) 
repeatedly, with the result
\begin{equation}
  G_{f,\beta} (x', x) = G^0_{f,\beta} (x', x) - \left( \sum_{n=1}^{\infty}
    2^{-n} \right) G^0_{f,\beta} (-x', x).
\end{equation}
This is indeed identical to (\ref{eq:G_free}).  It is clear that all terms in 
(\ref{eq:MRE_I}) are necessary to reproduce the correct result.  In addition, 
we cannot change the order of integration and taking the limit in 
(\ref{eq:MRE_I}).  In fact, since one has $\lim_{x''\downarrow 0} 
\partial_{x''} G^0_{f,\tau'}(0, x'') = 0$, this would give an incorrect 
result.  This suggests that (\ref{eq:MRE_I}) is not the most convenient form 
to use.

A slightly different series is obtained by modifying (\ref{eq:PDX_I}) as 
follows:
\begin{equation}
  \label{eq:PDX_II}
  G^c_\beta (x', x) = -\lim_{x''\uparrow\downarrow 0} \int_0^\beta
    \rmd\tau\,  \frac{\partial}{\partial x''} G_{\beta-\tau} (x', x'')\,
    G^0_\tau (0, x)
\end{equation}
where the limit $x\uparrow\downarrow 0$ is the average of the limits 
$x\uparrow 0$ and $x\downarrow 0$.  Since $G_\tau (x', x)$ vanishes for $x < 
0$, at least for a hard wall, we have merely added zero to the right-hand side 
of (\ref{eq:PDX_I}).  If we iterate (\ref{eq:PDX_II}), with (\ref{eq:split}) 
inserted, we get the resummed PDX series
\begin{eqnarray}
  \label{eq:MRE_II}
\fl  G_\beta (x', x) = G^0_\beta (x', x) 
- \lim_{x''\downarrow\uparrow 0}
  \int_0^\beta \rmd\tau\, 
  \frac{\partial}{\partial x''} G^0_{\beta-\tau} (x', x'')\, G^0_\tau (0, x)
  \nonumber\\
+ \lim_{x''\downarrow\uparrow 0}
  \int_0^\beta \rmd\tau\, 
  \frac{\partial}{\partial x''} \lim_{x'''\downarrow\uparrow 0}
  \int_0^{\beta-\tau} \rmd\tau'\nonumber\\
\times   \frac{\partial}{\partial x'''}
  G^0_{\beta-\tau-\tau'} (x', x''')\, G^0_{\tau'} (0, x'')\, G^0_\tau (0, x)
  -\ldots
\end{eqnarray}

Let us consider again the case $V_0(x) = 0$.  One easily verifies that the 
correction $G^c_\beta$ is given by the second term of (\ref{eq:MRE_II}) alone. 
The convergence of the resummed PDX series is thus found to be much better 
than that of of the original one. All higher-order terms in (\ref{eq:MRE_II}) 
vanish separately in the present case, since one may prove
\begin{equation}
  \lim_{x''\uparrow\downarrow 0} \frac{\partial}{\partial x''} 
  \lim_{x'''\uparrow\downarrow 0} \int_0^\tau \rmd\tau'
   \frac{\partial}{\partial x'''}
  G^0_{f,\tau - \tau'} (x', x''')\, G^0_{f,\tau'} (0, x'') = 0.
\end{equation}
Note that here we are allowed to interchange the order of integration and 
taking the limit. This property is an additional advantage of the series in 
(\ref{eq:MRE_II}). Returning to the general case with $V_0(x) \neq 0$, we 
expect that both favourable properties of the resummed PDX series (fast 
convergence and invariance under interchange of the order of integration and 
taking the limit) are conserved. Of course, in general the series will no 
longer terminate after the second term. Nevertheless, in some applications 
only a few terms in the expansion are relevant, as we shall see in the 
following.

The resummed PDX series (\ref{eq:MRE_II}) is of the general form
\begin{equation}
  \label{eq:MRE_III}
  G_\beta(x', x) = \sum_{n=0}^\infty G^{(n)}_\beta(x', x)
\end{equation}
where we put $G^{(0)}_\beta = G^0_\beta$.  The term of order $n$ involves $n$ 
positions at the boundary. It can be seen as arising from paths along which 
the particle hits the boundary $n$ times. These multiple reflections at the 
boundary form the basis of the multiple-reflection expansion, which was 
derived by Balian and Bloch \cite{BAB:1970} quite some time before the 
path-decomposition expansion was written down. A close inspection shows that 
the two expansions are completely equivalent. 

Note that in principle the PDX formula (\ref{eq:PDX_I}) and the PDX series 
(\ref{eq:MRE_I}) can be applied to any problem involving distinct spatial 
regions, for example to tunneling problems \cite{AUK:1985}. In contrast, the 
modified PDX formula (\ref{eq:PDX_II}) depends on the presence of a {\em hard} 
wall. The application of the resummed PDX series (\ref{eq:MRE_II}) is likewise 
limited to hard-wall problems only.

\section{Magnetic Field}
\label{sec:magnetic_field}

We will now apply the methods of the previous section to a confined 
free-electron gas in a uniform magnetic field. The Hamiltonian is given by
\begin{equation}
  H = \frac{1}{2} \left( \vect{p} - \vect{A} \right)^2 + V_w(\vect{r})
\end{equation}
where $V_w$ is again a hard-wall potential. Because of the symmetry of the 
problem we will choose the Landau gauge $\vect{A} = (0, Bx, 0)$. The factor 
$e/c$, with $e$ the charge of the particles, has been absorbed in the constant 
$B$.

The presence of the vector potential complicates matters.  The Feynman-Kac 
representation (\ref{eq:Feynman-Kac}) of the path integral is no longer valid. 
We have to use the Feynman-Kac-It\^o formula instead, which in the special 
case of $\nabla \cdot \vect{A} = 0$ reads \cite{ROEP:1996}
\begin{equation}
  \label{eq:Feynman-Kac-Ito}
\fl  G_\beta (\vect{r}', \vect{r}) = 
\int \rmd \mu_{\vect{r}, 0}^{\vect{r}', \beta}(\omega) \exp\left[
    -\int_0^\beta \rmd\tau\, V_w(\omega(\tau)) + \rmi \int_0^\beta
    \rmd\tau\, \dot\omega(\tau) \cdot \vect{A}(\omega(\tau))\right].
\end{equation}
If we let $\omega_x$ denote the $x$-component of the path, we can replace the 
exponential factor containing $V_w$ by $\theta (\inf_\tau \omega_x(\tau))$.  
Since the factor that contains the vector potential is independent of $z$, the 
integral over the $z$-component of the path gives a trivial factor 
$(2\pi\beta)^{-1/2} \exp [-(z'-z)^2/2\beta]$. The part of the Green function 
that depends on $x$ and $y$ will be denoted by $G_{\perp,\beta}(\vect{r}', 
\vect{r})$ in the following.

The path integral over the $y$-component of the path can be evaluated by a 
Fourier-transform technique. In fact, discretizing the $x$- and $y$-components 
of the path, with $n$ intermediate points, we write $\omega(\tau_m)=\vect{r}_m 
=(x_m,y_m) $, with $\vect{r}_0 = \vect{r}$ and $\vect{r}_{n+1} = \vect{r}'$.  
The integral in the exponent of (\ref{eq:Feynman-Kac-Ito}) is then given by 
$\int_0^\beta \rmd\tau\, \dot\omega(\tau) \cdot \vect{A}(\omega(\tau)) = 
\sum_{m=0}^{n} (\vect{r}_{m+1} - {\vect r}_m) \cdot \vect{A}(\vect{r}_m)$ (in 
It\^o's convention). We get
\begin{eqnarray}
\fl  G_{\perp,\beta}(\vect{r}', \vect{r}) = \prod_{m=1}^{n+1}
  \int \rmd \vect{r}_m\, (2\pi\epsilon_n)^{-1}
  \exp\left[-\frac{(\vect{r}_m-\vect{r}_{m-1})^2}{2\epsilon_n} \right] 
  \theta(x_m) \nonumber\\
\times \exp \left[ \rmi (y_m-y_{m-1}) B x_{m-1} \right] 
  \delta (\vect{r}_{n+1} - \vect{r}').
\end{eqnarray}
The integrals over $y_m$ can now be carried out by using the standard Fourier 
representation of the Dirac $\delta$-function. This introduces an integral 
over an additional variable $k$.  Going back to the continuum limit for the 
path integral over the $x$-component of the path we arrive at
\begin{equation}
  \label{eq:1D-PI}
  G_{\perp,\beta}(\vect{r}', \vect{r}) = (2\pi)^{-1}
  \int_{-\infty}^{\infty} \rmd k\, \rme^{\rmi k(y'-y)} \bar
  G_\beta(x'-k/B, x-k/B, k)
\end{equation}
with
\begin{equation}
  \label{eq:Gbar}
\fl  \bar G_\beta(x', x, k) = \int \rmd\mu_{x, 0}^{x', \beta}
  (\omega_x) \theta (\inf_\tau \omega_x(\tau) + k/B) \rme^{- \frac{1}{2}
  B^2 \int_0^\beta \rmd\tau\, \left[\omega_x(\tau)\right]^2}.
\end{equation}
This function $\bar G_\beta$ is the propagator for a particle in a 
one-dimensional harmonic potential with a wall at the position $-k/B$. 

We are now in a position to use the PDX techniques from the previous section. 
The leading term in the PDX series is found by omitting the wall. In that case 
the propagator $\bar G_\beta$ becomes \cite{ROEP:1996}:
\begin{equation}
  \label{eq:harmonic-oscillator}
\fl \bar G^0_\beta (x', x, k) = \left[ \frac{B}{2\pi\sinh(\beta B)}
  \right]^{1/2} \exp \left[ -\frac{B({x'}^2+x^2)}{2\tanh(\beta B)} +
  \frac{B x' x}{\sinh(\beta B)} \right].
\end{equation}
As a matter of fact, $\bar G^0_\beta$ is independent of $k$, since the only 
$k$-dependence in (\ref{eq:Gbar}) is in the position of the wall. After 
performing the integral over $k$, which is Gaussian, we find that the leading 
term in the PDX series is given by
\begin{equation}
\label{eq:first_reflection}
\fl  G^{(0)}_{\perp,\beta} (\vect{r}', \vect{r}) 
  = \frac{B}{4\pi\sinh(\beta B/2)} \exp \left
  [ -\frac{B}{4 \tanh (\beta B/2)} (\vect{r}' - \vect{r})^2
  + \frac{\rmi B}{2} (x' + x)(y' - y) \right]
\end{equation}
which is indeed the Green function in the Landau gauge for the unconfined 
system. 

The next term in the resummed PDX series (\ref{eq:MRE_II}) (or 
(\ref{eq:MRE_III})) is more complicated.  The integral over $k$ is again 
Gaussian (in fact it is Gaussian for all terms), but the additional integral 
over $\tau$ is not.  If we set $t_1 = \tanh(\tau B/2)$,  $s_1 = \sinh(\tau 
B/2)$, $t_2 = \tanh((\beta-\tau)B/2)$ and $s_2 = \sinh((\beta-\tau)B/2)$, we 
can write
\begin{equation}
\label{eq:second_reflection}
  G^{(1)}_{\perp,\beta} (\vect{r}', \vect{r}) = - \frac{B^2}{16 \pi^{3/2}}
  \int_0^\beta \rmd\tau f^{(1)}_{\beta, \tau} (\vect{r}', \vect{r}) \exp\left
  [ g^{(1)}_{\beta, \tau} (\vect{r}', \vect{r}) \right]
\end{equation}
with
\begin{equation}
  f^{(1)}_{\beta, \tau} (\vect{r}', \vect{r}) = \frac{1}{2} B^{1/2} \frac{(t_1
  t_2)^{1/2}}{s_1 s_2 (t_1 + t_2)^{1/2}} \left[\frac{x'}{t_1} +
  \frac{x}{t_2} + \rmi (y' - y)\right]
\end{equation}
and
\begin{equation}
\fl  g^{(1)}_{\beta, \tau} (\vect{r}', \vect{r}) = \frac{B}{4} 
  \left\{ \frac{[x' t_1 + x t_2 + \rmi (y' - y)]^2}{t_1 + t_2}
    - \left(t_1 + \frac{1}{t_1}\right) {x'}^2
    - \left(t_2 + \frac{1}{t_2}\right) x^2
  \right\}.
\end{equation}
Similar expressions can be found in \cite{JOS:1995}. Note that the formulas in 
\cite{JOS:1995} slightly differ from those given above. We have made use of 
the property $G_\beta (\vect{r}', \vect{r}) = [G_\beta (\vect{r}, 
\vect{r}')]^*$ and of the possibility to change $\tau$ into $\beta-\tau$ to 
write  $f^{(1)}$ and $g^{(1)}$ in a form that is more symmetric.

The higher-order terms in the resummed PDX series can be found along similar 
lines. For the special case $\vect{r}'=\vect{r}$ they have been collected in 
\ref{app:higher_reflections}. They are found to agree with those derived in 
\cite{JOS:1995}, after appropriate symmetrization. 

\section{Asymptotics (non-degenerate case)}
\label{sec:non-degenerate}

The particle density and the (electric) current density can both be found from 
the Green function.  In the absence of quantum degeneracy the particle density 
is directly related to $G_{\perp,\beta}$ via
\begin{equation}
  \label{eq:rho}
  \rho(x) = \frac{\rho_0}{Z_\perp} G_{\perp,\beta}(\vect{r}, \vect{r})
\end{equation}
where $\rho_0$ is the bulk density and $Z_\perp = B/[4\pi\sinh(\beta B/2)]$ is 
the transverse one-particle partition function per unit area for the bulk.  
The expression for the current density is slightly more complicated, involving 
derivatives of $G_{\perp,\beta}$:
\begin{equation}
  \label{eq:j_y}
  j_y(x) = \frac{\rho_0}{Z_\perp}
  \frac{1}{2\rmi}\left[\frac{\partial}{\partial y'}
    G_{\perp,\beta}(\vect{r}', \vect{r})
    - \frac{\partial}{\partial y} G_{\perp,\beta} (\vect{r}',
  \vect{r})\right]_{\vect{r}'=\vect{r}} -  Bx \rho(x).
\end{equation}

Using only the $n=0$ term of the PDX series in the expression for $\rho(x)$ 
yields the bulk density $\rho_0$, as it should, since $G^0_{\perp, 
\beta}(\vect{r}, \vect{r})=Z_\perp$.  Therefore we will consider the excess 
particle density $\delta\rho(x) = \rho(x)-\rho_0$ instead of $\rho(x)$ in the 
following.  Since there is no bulk current, the $n=0$ term of the PDX series 
does not contribute to the current density.

To determine the profiles of the excess particle density and the current 
density for arbitrary distances from the wall we need to evaluate all terms in 
the resummed PDX series. However, the $\tau$-integral in 
(\ref{eq:second_reflection}) cannot be carried out analytically. Likewise, 
evaluation of the multiple $\tau$-integrals in the higher-order terms given in 
\ref{app:higher_reflections} is in general not possible. 

For large distances from the wall (in units of the magnetic length 
$1/\sqrt{B}$)  the leading contribution to the profiles comes from the $n=1$ 
term in the resummed PDX series, as we will discuss presently. Moreover, the 
$\tau$-integral in (\ref{eq:second_reflection}) can be evaluated analytically 
in that limit. It is thus possible to derive asymptotic expressions for the 
profiles of the excess particle density and the current density that are valid 
far large $\sqrt{B}\,x$.

A change of variables $y = z^2 / ({z_0}^2 - z^2)$, with $z = \tanh[B (2\tau - 
\beta) / 4]$ and $z_0 = \tanh(B \beta / 4)$, brings 
(\ref{eq:second_reflection}) into the form
\begin{eqnarray}
\label{eq:G_1}
\fl  G^{(1)}_{\perp,\beta} (\vect{r}, \vect{r}) = - \frac{B^{3/2} x}{8 \sqrt{2}
      \pi^{3/2}} \frac{1 - {z_0}^2}{z_0^{3/2}} \exp\left[-\frac{B
      x^2}{2 z_0}\right] \int_0^\infty \frac{\rmd y}{\sqrt{y(1+y)}}
      \sqrt{1 + (1-{z_0}^2)y} \nonumber\\
   \times \exp\left[-\frac{(1-{z_0}^2)y}{2 z_0}Bx^2 \right].
\end{eqnarray}
Because of the presence of $Bx^2$ in the exponent, only small values of 
$(1-{z_0}^2) y / z_0$ contribute to the integral for large $\sqrt{B}\, x$.  
Since one has $0 \leq z_0 < 1$, this implies small values of $(1-{z_0}^2) y$. 
Note that this does not necessarily mean that $y$ itself is small, as $z_0$ 
may be close to 1. For large $\sqrt{B}\, x$ the factor $\sqrt{1 + (1-{z_0}^2) 
y}$ in the integrand can be replaced by 1.  Subsequently, we can use
\begin{equation}
  \int_0^\infty \frac{\rmd y}{\sqrt{y(1+y)}} \rme^{-a y} = \rme^{a/2} K_0(a/2)
\end{equation}
where $K_0$ is the modified Bessel function of the second kind.  In this way 
we arrive at the following asymptotic expression for the transverse part of 
the Green function for large $\sqrt{B}\,x$:
\begin{equation}
  \label{eq:G_MB}
\fl  G^{(1)}_{\perp,\beta} (\vect{r}, \vect{r}) \approx - \frac{B^{3/2} x}{8 \sqrt{2} \pi^{3/2}}
  \frac{1-{z_0}^2}{{z_0}^{3/2}} \exp\left[ - \frac{1+{z_0}^2}{4 z_0}
  Bx^2\right] K_0 \left(\frac{1-{z_0}^2}{4 z_0} Bx^2\right).
\end{equation}

The next term in the asymptotic expansion of $G^{(1)}_{\perp,\beta}$ is
\begin{eqnarray}
\label{eq:nextorder}
- \frac{B^{3/2} x}{8 \sqrt{2} \pi^{3/2}}
  \frac{(1-{z_0}^2)^2}{4 {z_0}^{3/2}} \exp\left( - \frac{1+{z_0}^2}{4 z_0}
  Bx^2\right) \nonumber\\
\times\left\{ K_1 \left(\frac{1-{z_0}^2}{4 z_0} Bx^2\right) -
  K_0 \left(\frac{1-{z_0}^2}{4 z_0} Bx^2\right) \right\}
\end{eqnarray}
which can be derived by substituting $\sqrt{1+(1-{z_0}^2) y} \approx 1 + 
(1-{z_0}^2)y/2$.  Since $z [K_1(z) - K_0(z)]/ K_0 (z)$ is bounded for all 
positive $z$, we see that (\ref{eq:nextorder}) is indeed of higher order in 
$1/(\sqrt{B}\,x)$. 

\begin{figure}
  \begin{center}
    \includegraphics[height=6cm]{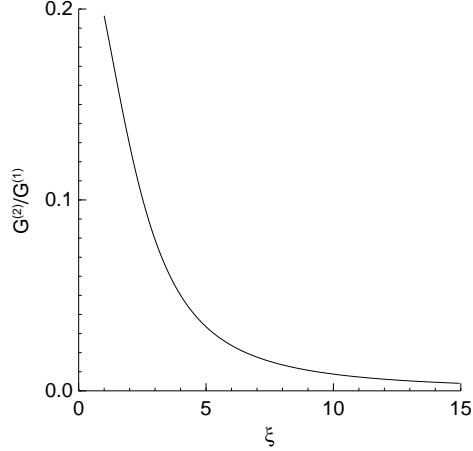}
  \end{center}
  \caption{Numerical results for
    $G^{(2)}_{\perp,\beta}/G^{(1)}_{\perp,\beta}$ as a function of 
    $\xi=\sqrt{B}\,x$, for $\beta B = 4$.}
  \label{fig:G2_G1}
\end{figure}

Having investigated the $n=1$ term in the resummed PDX series, we may turn to 
the higher orders. From a detailed analysis (see \ref{app:higher_reflections}) 
it is found that all terms with $n > 1$ are of higher order in $1/(\sqrt{B}\,x)$ 
in comparison with (\ref{eq:G_MB}).  In figure~\ref{fig:G2_G1} we have plotted 
$G^{(2)}_{\perp,\beta}/G^{(1)}_{\perp,\beta}$ as a function of 
$\xi=\sqrt{B}\,x$, for a representative value of $\beta B$.
The decay is in good agreement with (\ref{eq:higher_reflections}).

The asymptotic expression for the excess particle density at large $\sqrt{B}\, 
x$ follows by substituting (\ref{eq:G_MB}) in (\ref{eq:rho}):
\begin{equation}
  \label{eq:rho_MB}
\fl  \delta\rho(x) \approx -\rho_0 \frac{B^{1/2} x}{\sqrt{2\pi}}
  {z_0}^{-1/2} \exp\left(-\frac{1+{z_0}^2}{4 z_0} Bx^2\right) K_0
  \left( \frac{1 - {z_0}^2}{4 z_0} Bx^2 \right)
\end{equation}
where we have used that $Z_\perp$ is given by $Z_\perp = B(1-{z_0}^2)/(8\pi 
z_0)$ in terms of $z_0$. In a similar way an asymptotic expression for the 
current density at large $\sqrt{B}\, x$ can be derived. In leading order it is 
found to be proportional to the asymptotic excess particle density:
\begin{equation}
  \label{eq:j_y_MB}
  j_y(x) \approx - \frac{1}{2}Bx \;\delta\rho(x)
\end{equation}
with $\delta\rho(x)$ given in (\ref{eq:rho_MB}).  This simple proportionality 
ceases to be valid, if higher-order terms are incorporated in the asymptotic 
expansion. Comparing (\ref{eq:j_y_MB}) to (\ref{eq:j_y}) we see that there is 
a compensation between the term proportional to $\rho(x)$ and the term that 
contains the derivatives of the Green function. For the $n=0$ contribution 
this compensation is complete, but for $n=1$ only half of the second term in 
(\ref{eq:j_y}) is cancelled, at least in leading order in $1/(\sqrt{B}\,x)$.

It must be stressed that both (\ref{eq:rho_MB}) and (\ref{eq:j_y_MB}) are 
valid for large $\sqrt{B}\,x$, whereas $\beta B$ may take arbitrary values. If, 
apart from $B x^2$, also $[(1-{z_0}^2)/z_0]Bx^2$ is large, we can simplify 
(\ref{eq:G_MB}) to
\begin{equation}
  \label{eq:G1-simplified}
  G^{(1)}_{\perp,\beta}(\vect{r},\vect{r}) \approx - \frac{B}{8\pi} 
\frac{\sqrt{1-{z_0}^2}}{z_0}  \exp\left(-\frac{Bx^2}{2z_0}\right)
\end{equation}
by using the asymptotic expansion for the modified Bessel function. In that 
case the excess particle density profile is asymptotically given by
\begin{equation}
  \label{eq:rho_MB-simplified}
  \delta \rho(x) \approx - \rho_0 \cosh(\beta B/4)
  \exp \left[ -\frac{Bx^2}{2 \tanh(\beta B/4)} \right].
\end{equation}
Large $[(1-{z_0}^2)/z_0]Bx^2$ implies that the regime $z_0 \rightarrow 1$ or 
$\beta B\rightarrow\infty$ is not included, whereas no such limitation is 
imposed on the use of (\ref{eq:rho_MB}).  For fixed $B$ this is not a serious 
limitation in the present context of a non-degenerate electron gas, since for 
$\beta\rightarrow\infty$ we have to use Fermi-Dirac statistics anyway. In the 
next section it will be shown that the Green function in the form 
(\ref{eq:G_MB}) is crucial to obtain information on the asymptotic profiles 
for a degenerate electron gas with Fermi-Dirac statistics.

\begin{figure}
  \begin{center}
    \includegraphics[height=6cm]{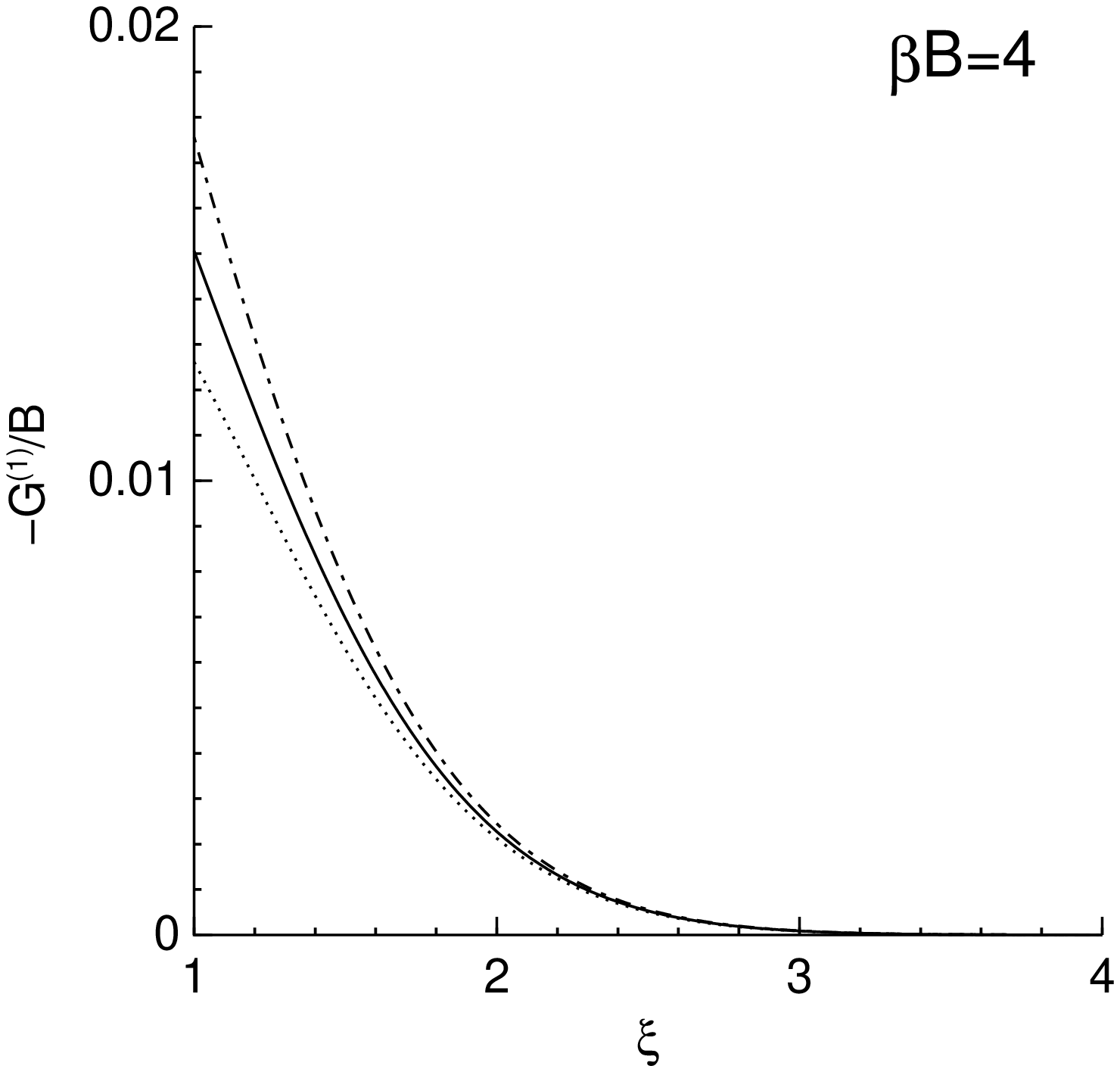}
    \includegraphics[height=6cm]{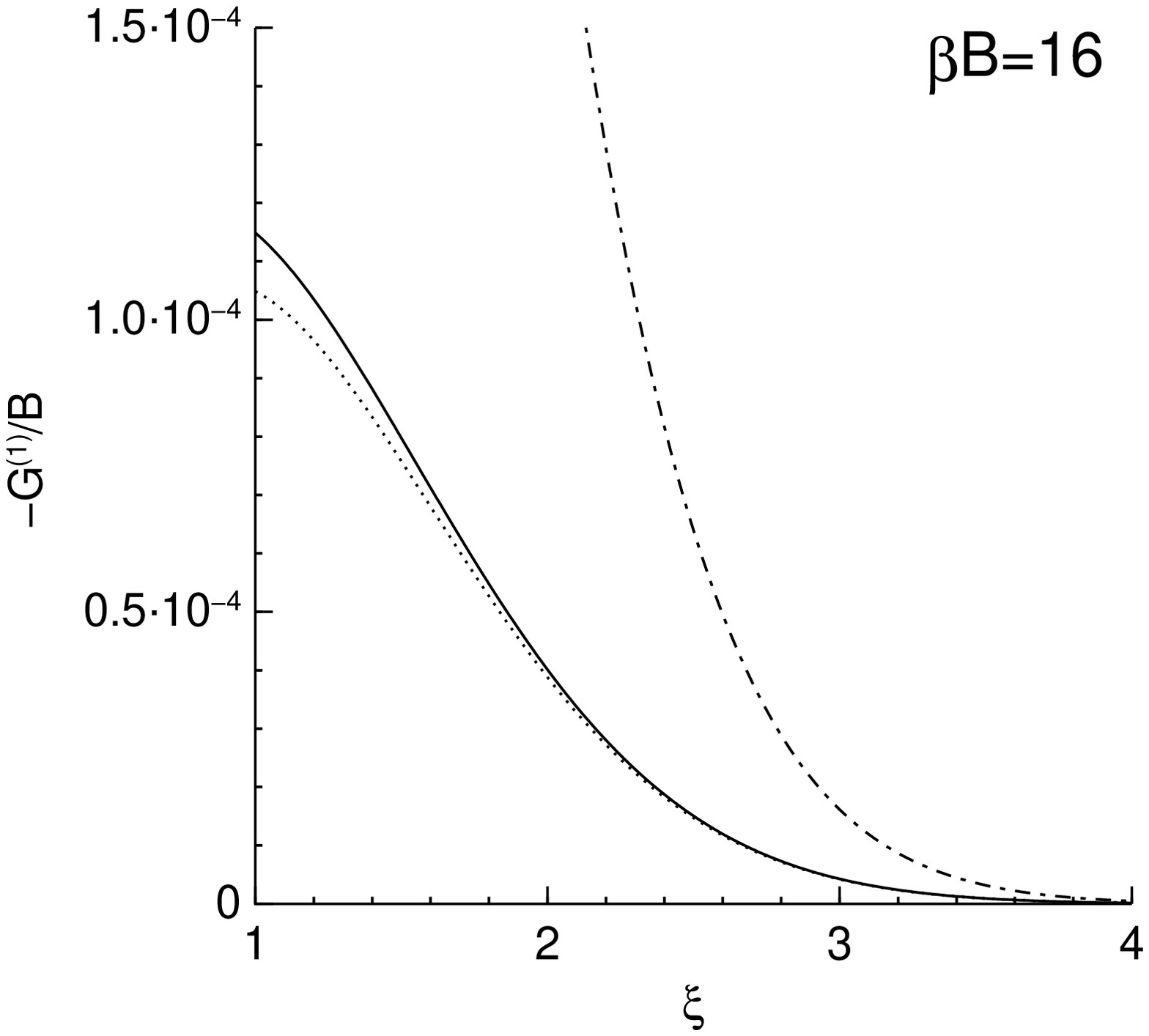}
  \end{center}
  \caption{Comparison between results from numerical integration of 
(\ref{eq:second_reflection}) or (\ref{eq:G_1}) (\full) and the asymptotic 
expressions (\ref{eq:G_MB}) (\dotted) and (\ref{eq:G1-simplified}) (\chain) for 
$-G^{(1)}_{\perp,\beta}/B$ as a function of $\xi$.}
  \label{fig:G1}
\end{figure}

To investigate the validity of the asymptotic expressions for $G^{(1)}$ 
mentioned above we have compared (\ref{eq:G_MB}) and (\ref{eq:G1-simplified}) 
with numerical results based on (\ref{eq:second_reflection}) or 
(\ref{eq:G_1}). The results are drawn in figure~\ref{fig:G1}.
It is clear that for $\beta B = 4$ both asymptotic expressions are adequate, 
even for relatively small values of $\xi$, whereas for $\beta B = 16$ 
the performance of (\ref{eq:G_MB}) is much better than that of 
(\ref{eq:G1-simplified}).

Finally, for small $B$, the expression (\ref{eq:rho_MB-simplified}) for the 
excess particle density yields
\begin{equation}
  \rho (x) \approx - \rho_0 \left( 1 + \frac{1}{32} \beta^2 B^2 -
  \frac{1}{24} \beta B^2 x^2 \right) e^{-2x^2/\beta}.
\end{equation}
Indeed these terms correspond to the leading terms (for large $x^2/\beta$) in 
the expression for the density up to order $B^2$, as given in \cite{JOS:1995}.

In closing this section on the non-degenerate electron gas we remark that the 
path-integral representation can be used to derive a strict bound on the 
particle density for all values of $x$. Upon setting $k' = x - k/B$ we find 
from (\ref{eq:1D-PI}), (\ref{eq:Gbar}) and (\ref{eq:rho}):
\begin{equation}
\fl  \delta\rho(x) = - \frac{\rho_0 B}{2 \pi Z_\perp}
  \int_{-\infty}^{\infty} \rmd k' \int \rmd
  \mu_{k',0}^{k',\beta}(\omega_x) \theta (k' - x - \inf_\tau
  \omega_x(\tau)) \rme^{-\frac{1}{2} B^2 \int_0^\beta
  \rmd\tau[\omega_x(\tau)]^2}.
\end{equation}
All paths that contribute to this integral must pass below the point $k' - x$, 
while starting and finishing at $k'$.  Now look at paths that go via a point 
below $k' - x$ precisely at $\tau = \beta/2$.  Since these form a subclass of 
all allowed paths, the corresponding path integral provides a lower bound on 
$\abs{\delta\rho(x)}$:
\begin{eqnarray}
\fl  \abs{\delta\rho(x)} \geq \frac{\rho_0 B}{2\pi Z_\perp}
  \int_{-\infty}^{\infty} \rmd k' \int_{-\infty}^{k'-x} \rmd x' 
\int \rmd\mu_{k',0}^{x',\beta/2}(\omega_x) \rme^{-\frac{1}{2} B^2 
\int_0^{\beta/2} \rmd\tau [\omega_x(\tau)]^2} \nonumber\\
\times\int \rmd\mu_{x',\beta/2}^{k', \beta}  (\omega_x) 
\rme^{-\frac{1}{2}B^2 \int_{\beta/2}^\beta \rmd\tau [\omega_x(\tau)]^2}.
\end{eqnarray}
The path integrals in this expression are now unrestricted, so that they are 
given by $\bar G^0_\beta$ (see (\ref{eq:harmonic-oscillator})).  Integration 
over $k'$ and $x' -k'$ (in that order) gives
\begin{equation}
  \abs{\delta\rho(x)} \geq \frac{1}{2} \rho_0 \Erfc \left(x
  \left[ \frac{B}{2 \tanh{\beta B/4}} \right]^{1/2} \right)
\end{equation}
which is the bound for all $x$ that we set out to derive. In the limit of 
large $x$ this implies
\begin{equation}
  \lim_{x\rightarrow\infty} x \exp \left[ \frac{Bx^2}{2\tanh(\beta B/4)}
  \right]\abs{\delta\rho(x)} \geq \frac{\rho_0}{2 \sqrt
  { \pi}}\left [ \frac{2 \tanh(\beta B/4)}{B} \right]^{1/2}.
\end{equation}
This inequality is consistent with (\ref{eq:rho_MB-simplified}), as it should 
be. In particular, the Gaussian decay of $\delta\rho(x)$, with the same 
characteristic length as in (\ref{eq:rho_MB-simplified}), is corroborated. 

As can be seen from the results (\ref{eq:rho_MB}) and (\ref{eq:j_y_MB}) the 
decay towards the bulk value of both the excess particle density and the 
current density is Gaussian, modulated by a Bessel function and an algebraic 
factor. For not too large $\beta B$ the decay of the excess particle density, 
as given by (\ref{eq:rho_MB-simplified}), is strictly Gaussian far from the 
edge. The asymptotic decay of the current density is likewise Gaussian, albeit 
with an extra algebraic factor. The characteristic length on which the 
Gaussian decay manifests itself is proportional to $[\tanh(\beta 
B/4)/B]^{1/2}$. As we have shown above, the Gaussian decay for the excess 
particle density is consistent with a lower bound that can be derived exactly. 
For the current density it is consistent with the upper bound on the absolute 
value of the current density that has been derived by Macris et al.\ 
\cite{MMP:1988}. However, it should be remarked that the upper bound obtained 
in that paper is rather wide. In fact, the characteristic length of the 
Gaussian function in their upper bound is the thermal wave length, which is 
independent of the magnetic field. This characteristic length is larger than 
that in the Gaussian found here, at least for non-vanishing magnetic fields.

\section{Asymptotics (degenerate case)}
\label{sec:degenerate}

The particle density ${\rho(x)}_{FD}$ of a degenerate Fermi-Dirac system at 
temperature $T=0$ and chemical potential $\mu$ is related to the density of 
the non-degenerate system by a Laplace transformation:
\begin{equation}
   \label{eq:laplace}
  \int_0^\infty \rmd\mu\,\rme^{-\beta\mu} {\rho(x)}_{FD} =
  \frac{2 Z}{\rho_0\beta} \rho(x).
\end{equation}
Here $Z = Z_\parallel Z_\perp$ is the total one-particle partition function 
per unit volume for the bulk, with $Z_\parallel = (2\pi\beta)^{-1/2}$; the 
factor $2$ takes the spin degeneracy into account. The relation 
(\ref{eq:laplace}) implies that we can calculate the excess particle density 
$\delta{\rho(x)}_{FD}$ from $\delta\rho(x)$ via an inverse Laplace transform 
\cite{SOW:1951}:
\begin{equation}
  \label{eq:rho_FD}
  \delta{\rho(x)}_{FD} = \frac{1}{2\pi \rmi} \int_{c-i\infty}^{c+i\infty}
  \rmd\beta\; \rme^{\beta\mu} \frac{2Z}{\rho_0\beta} \;\delta\rho(x)
\end{equation}
with arbitrary $c > 0$. Hence, the asymptotic behaviour of the excess particle 
density of the degenerate system for large $\sqrt{B}\,x$ can be obtained on the 
basis of the results of the previous section.   

Let $\xi = \sqrt{B}\, x$ and $\nu = \mu /B$, and introduce a new integration 
variable $t$ by writing $\beta=\xi(\rmi t+1)/B$, so that $c$ is given by 
$\xi/B$.  If we express the right-hand side of (\ref{eq:rho_MB}) in the 
variables $\xi$ and $\nu$, and substitute it into (\ref{eq:rho_FD}), we get
\begin{eqnarray}
\label{eq:rho_FDa}
  \delta{\rho(x)}_{FD} \approx - \frac{B^{3/2}\xi^{1/2}}{16 \pi^3}
  \int_{-\infty}^{\infty} \rmd t
  \frac{\rme^{\nu\xi(\rmi t+1)}}{(\rmi t+1)^{3/2}}
  \frac{1-{z_0}^2}{{z_0}^{3/2}} \nonumber\\
\times \exp \left(-\frac{1+{z_0}^2}{4z_0}
  \xi^2\right) K_0\left(\frac{1-{z_0}^2}{4 z_0} \xi^2\right).
\end{eqnarray}
In the new variables we have $z_0 = \tanh[\xi(\rmi t+1)/4]$, so that large 
$\xi$ implies $z_0 \approx 1$ and $1-{z_0}^2 \approx 4 \exp[-\xi(\rmi 
t+1)/2]$. Consequently, the argument of $K_0$ in (\ref{eq:rho_FDa}) is small 
in absolute value, so that we can use the series representation
\begin{equation}
  K_0(z) = \sum_{n=0}^\infty \left[\sum_{m=1}^n \frac{1}{m} - \gamma -
    \log \left(\frac{z}{2}\right) \right] \frac{1}{2^{2n}(n!)^2} z^{2n}
\end{equation}
for the modified Bessel function. In this way we get 
\begin{eqnarray}
\label{eq:rho_FDb}
\delta{\rho(x)}_{FD} \approx -\frac{B^{3/2}\xi^{1/2}}{4\pi^3}
  \rme^{-\xi^2/2} \sum_{n=0}^{\infty} 
\frac{\xi^{4n}}{2^{2n}(n!)^2} \int_{-\infty}^{\infty} \rmd t \nonumber\\
\times   \left[ (\rmi t + 1)\frac{\xi}{2} + \sum_{m=1}^{n}
  \frac{1}{m}- \gamma - \log\left(\frac{\xi^2}{2}\right)  \right]
  \frac{\rme^{[\nu-(n+1/2)] \xi(\rmi t+1)}}{(\rmi t+1)^{3/2}}.
\end{eqnarray}
Upon using the identity \cite{ERD:1954}
\begin{equation}
  \int_{-\infty}^\infty \rmd t \frac{\rme^{(\rmi t+1)x}}{(\rmi t+1)^\nu} = 
\theta(x)  \frac{2\pi x^{\nu-1}}{\Gamma(\nu)} \qquad (\nu > 0)
\end{equation}
we arrive at the asymptotic expression for the excess particle density
\begin{eqnarray}
  \label{eq:rho_FDc}   
\fl  \delta{\rho(x)}_{FD} \approx - \frac{B^{3/2}\xi}{\pi^{5/2}}
  \rme^{-\xi^2/2} \sum_n^{[\nu-1/2]} \frac{\xi^{4n}}{2^{2n}(n!)^2} 
\left[\nu -  \left(n+\frac{1}{2}\right)\right]^{1/2} \nonumber\\
\times  \left[ \frac{1}{4[\nu-(n+\frac{1}{2})]} +
  \sum_{m=1}^n \frac{1}{m} - \gamma - \log \left(\frac{\xi^2}{2}\right) \right]
\end{eqnarray}       
with $[x]$ the largest integer less than or equal to $x$. The asymptotic 
expression derived here is valid for large $\xi$ and fixed $\nu$. 

The profile of the current density for large $\xi$ and fixed $\nu$ can 
likewise be obtained from the results of the previous section. In fact, 
because of the linearity of the inverse Laplace transform the asymptotic form 
of the current density is related to that of the excess particle density in 
the same way as in (\ref{eq:j_y_MB}):
\begin{equation}
  \label{eq:j_y_FD}
  {j_y(x)}_{FD} \approx - \frac{1}{2} B x \; \delta {\rho(x)}_{FD}.
\end{equation}

The expressions (\ref{eq:rho_FDc}) and (\ref{eq:j_y_FD}) for the asymptotic 
profiles of the excess particle density and the current density are identical 
to the leading terms of the asymptotic expansions in \cite{KES:1998}, which 
have been obtained by solving the eigenvalue problem and analysing the 
asymptotics of the eigenfunctions. It is also possible to recover the 
higher-order terms of \cite{KES:1998} by inserting higher-order terms in the 
approximate expressions for the factors $z_0$ and $1-z_0^2$ in the integrand 
in (\ref{eq:rho_FDa}), taking into account corrections like 
(\ref{eq:nextorder}), and including more terms in the resummed PDX series as 
well.

The asymptotic behaviour of (\ref{eq:rho_FDc}) (and of (\ref{eq:j_y_FD})) is 
Gaussian in $\xi$, so that the characteristic length is the magnetic length 
$1/\sqrt{B}$ for a completely degenerate electron gas. Furthermore, the 
Gaussian is multiplied by a prefactor that depends algebraically and 
logarithmically on $\xi$. For $\nu$ just above a half-odd integer, that is, 
for chemical potentials $\mu$ slightly above a Landau level, the profile of 
the excess particle density shows a singular behaviour that is a remnant of 
the de Haas-van Alphen effect. A numerical assessment of the convergence of 
this  asymptotic expression can be found in \cite{KES:1998}.

The dominant term in the asymptotic behaviour comes from the highest Landau 
level with the label $[\nu-1/2]$.  Since the prefactor of the Gaussian in this 
term is proportional to $\xi^{4[\nu-1/2]+1}$, the onset of the Gaussian decay 
shifts to larger and larger values of $\xi$, if $\nu$ increases. In fact, 
(\ref{eq:rho_FDc}) is useful only for $\xi^2 \gg \nu$, or equivalently for $x$ 
large compared to the cyclotron radius  $\sqrt{\mu}/B$ of particles at the 
Fermi level. If $\nu$ is large, a different behaviour can be expected in the 
regime $\xi^2 \approx \nu$, before the ultimate Gaussian decay sets in at 
$\xi^2 \gg \nu$.

In conclusion, we have studied the edge effects in the excess particle density 
and the current density of a magnetized free-electron gas, which is confined 
by a hard wall. In particular, we have studied the long-range influence of the 
wall on these quantities, by determining their asymptotic profiles both for 
the non-degenerate case and for strong degeneracy. New results have been 
obtained for both these cases. In the former case the asymptotic spatial 
profiles were found to be Gaussian (or Gaussian modulated by a Bessel 
function), with a characteristic length that is proportional to $[\tanh(\beta 
B/4)/B]^{1/2}$. In the latter case the asymptotic behaviour depends on the 
number of filled Landau levels $n=[\mu/B-1/2]$. In fact, it is determined by a 
Gaussian, with a characteristic length equal to the magnetic length 
$1/\sqrt{B}$, multiplied by a polynomial and a logarithmic prefactor. Since 
the degree of the polynomial prefactor grows with $n$, the Gaussian 
character of the asymptotics comes to the fore only for distances that are 
large compared to $\sqrt{n}$ times the magnetic length.

\ack

This investigation is part of the research programme of the `Stichting voor 
Fundamenteel Onderzoek der Materie (FOM)', which is financially supported by 
the `Nederlandse Organisatie voor Wetenschappelijk Onderzoek (NWO)'.

\appendix

\section{Higher-order terms in the PDX series}
\label{app:higher_reflections}

In this appendix we study the asymptotic behaviour of the terms with $n>1$ in 
the resummed PDX series for the Green function, for large values of 
$\sqrt{B}\,x$. The general form of the term of order $n$ in the resummed PDX 
series is
\begin{eqnarray}
\label{eq:highorder}
\fl  G^{(n)}_{\perp,\beta} (\vect{r}', \vect{r}) = (-)^n
  \frac{B^{n+1}}{2^{n+3} \pi^{(n+2)/2}} \int_0^\beta \rmd\tau_1 \cdots
  \int_0^\beta \rmd\tau_n\; \theta(\tau_{n+1}) \nonumber\\
\times f^{(n)}_{\beta,\tau_1,\ldots,\tau_n} (\vect{r}', \vect{r})
  \exp[g^{(n)}_{\beta, \tau_1, \ldots, \tau_n} (\vect{r}', \vect{r})]
\end{eqnarray}
with $\tau_{n+1} = \beta - \sum_{i=1}^n \tau_i$. The functions $f^{(n)}$ and 
$g^{(n)}$ can be found in \cite{JOS:1995}. Here we collect them for the case 
$\vect{r}'=\vect{r}$, which is relevant for the particle density.  In that 
case we can symmetrize the expressions in $t_1$ and $t_2$. As a result they 
get the form
\begin{eqnarray}
\fl  f^{(n)}_{\beta, \tau_1, \ldots, \tau_n} (\vect{r}, \vect{r}) =
  \left(\prod_{i=1}^{n+1}\frac{{t_i}^{3/2}}{s_i}\right)
  \left(\sum_{i=1}^{n+1} t_i\right)^{-(n+1)/2} \frac{t_1 +
  t_2}{2 t_1 t_2}
\left\{ 2\;\left(\frac{1}{2}\right)_{n/2}\;
  \delta_{n,\mathrm{even}} \right.\nonumber\\
 + \sum_{p=0}^{[(n-1)/2]}  \frac{(1/2)_p}{2^{n-2p-1}} 
\left(\sum_{i=1}^{n+1} t_i\right)^{p-n/2} (Bx^2)^{n/2-p} (t_1 + t_2)^{n-2p-1}
\nonumber\\
\times  \left[(t_1 + t_2) \binom{n-1}{2p-1} - \left(\sum_{i=3}^{n+1}
  t_i\right) \binom{n-1}{2p}\right. \nonumber\\
 \left.\left. + \frac{t_1 + t_2}{t_1 t_2}
  \binom{n-1}{2p} + \frac{1}{t_1 t_2} \left( \sum_{i=3}^{n+1} t_i \right)
  \binom{n-1}{2p} \right] \right\} 
\end{eqnarray}   
and
\begin{equation}
  g^{(n)}_{\beta, \tau_1, \ldots, \tau_n} (\vect{r}, \vect{r}) = - \frac{Bx^2}{4} (t_1 + t_2) \left[ 1 +
  \frac{1}{t_1 t_2} - \frac{t_1 + t_2}{\sum_{i=1}^{n+1} t_i} \right]
\end{equation}
with $t_i = \tanh(B\tau_i/2)$, $s_i = \sinh(B\tau_i/2)$ and $(a)_n$ 
Pochhammer's symbol $a(a+1)\cdots(a+n-1)$.

For large $\sqrt{B}\,x$ the dominant contribution to the integral
comes from the
integration region for which the factor multiplying $Bx^2$ in $|g^{(n)}|$ is 
minimal. This is the case for $\tau_1=\tau_2=\beta/2$ and $\tau_i=0$  (with 
$3\leq i\leq n+1$). Therefore, we introduce on a par with $z_0 = \tanh 
(B\beta/4)$ the new integration variables $z_+ = \tanh[B(\beta - \tau_1 - 
\tau_2)/4]$, $z_- = \tanh[B(\tau_1-\tau_2)/4]$ and for $n>2$ also $z_i = 
\tanh(B\tau_i/2) = t_i$ ($i=3,\ldots,n$). If the integrations are carried out 
in the order $z_i, z_+$ and $z_-$, the allowed intervals of these variables 
are $z_- \in [-z_0, z_0]$,  $z_+ \in [0,(z_0-|z_-|)/(1-z_0|z_-|)]$,  and $z_i 
\in [0,2z_+/(1+z_+^2)]$, with an additional condition on $z_i$ resulting from 
the $\theta$-function in (\ref{eq:highorder}). 

We now have to rewrite the integrand of (\ref{eq:highorder}) in terms of $z_+, 
z_-$ and $z_i$. Let us consider small values of $z_+$ and $z_i$. The function 
$g^{(n)}$ then gets the form
\begin{eqnarray}
\fl  g^{(n)}_{\beta, \tau_1, \ldots, \tau_n} (\vect{r}, \vect{r}) \approx - \frac{Bx^2}{2}
  \left[ \frac{z_0(1-{z_-}^2)}{{z_0}^2-{z_-}^2} \right.\nonumber\\
\left.+ \frac{{z_0}^4{z_-}^2
  + {z_0}^2{z_-}^4 -4{z_0}^2{z_-}^2 + {z_0}^2 + {z_-}^2}{({z_0}^2 -
  {z_-}^2)^2} z_+ + \ldots \right].
\end{eqnarray}
{}From the right-hand side it is seen that it is indeed true that only small 
values of $z_+$ contribute to the integral in (\ref{eq:highorder}), as $Bx^2$ 
is large. In turn this implies that all $z_i$ have to be small as well, 
whereas no condition of smallness is imposed on $z_-$. In $f^{(n)}$ only the 
$p=0$ terms are relevant for large $Bx^2$, since these give the terms with the 
highest power of $x$. As a consequence we can write $f^{(n)}$ as
\begin{eqnarray}
\fl  f^{(n)}_{\beta, \tau_1, \ldots, \tau_n} (\vect{r},\vect{r}) \approx (Bx^2)^{n/2} \frac{{z_0}^{1/2}
  (1-{z_0}^2)}{2^{n-1/2}}
  \left(\frac{1-{z_-}^2}{{z_0}^2-{z_-}^2}\right)^{3/2} \nonumber\\
\times \left(2z_+ -
  \sum_{i=3}^{n} z_i\right)^{1/2} \left(\prod_{i=3}^{n} z_i^{1/2}\right)
\end{eqnarray}
again for small values of $z_+$ and $z_i$. Finally, the $\theta$-function in 
(\ref{eq:highorder}) is equal to $\theta(2z_+-\sum_{i=3}^n z_i)$ in the 
neighbourhood of $z_+=z_i=0$.

Since in the approximation considered here $g^{(n)}$  does not depend on $z_i$ 
anymore, the integral over these variables can be evaluated easily:
\begin{eqnarray} 
\fl \left(\prod_{i=3}^{n} \int_0^{2z_+/(1+{z_+}^2)} \rmd z_i \right)
  \theta\left(2z_+ - \sum_{i=3}^{n} z_i \right) \left(2z_+ -
    \sum_{i=3}^{n} z_i\right)^{1/2} \left(\prod_{i=3}^{n}
    {z_i}^{1/2}\right)\nonumber\\
= \frac{\pi^{(n-1)/2}}{2^{n-1} \Gamma[3(n-1)/2]} (2z_+)^{(3n-5)/2}.
\end{eqnarray}
To calculate the integral we have extended the upper limit to $\infty$, since 
the condition $z_+ \ll 1$ guarantees that only small values of $z_i$ 
contribute anyway. The subsequent integral over $z_+$ gets the following form:
\begin{eqnarray}
  \int_0^{\frac{z_0-\abs{z_-}}{1-z_0\abs{z_-}}} \rmd z_+\;
  (2z_+)^{(3n-5)/2} \nonumber\\
\times\exp\left[-\frac{Bx^2}{2} \frac{{z_0}^4{z_-}^2
  + {z_0}^2{z_-}^4 -4{z_0}^2{z_-}^2 + {z_0}^2 + {z_-}^2}{({z_0}^2 -
  {z_-}^2)^2} z_+ \right].
\end{eqnarray}
Again we can choose $\infty$ for the upper limit, since only small values of 
$z_+$  are significant; the integral can then be carried out trivially. 

We are left with the integral over $z_-$. Leaving it in its original form we 
arrive at
\begin{eqnarray}
\fl  G^{(n)}_{\perp,\beta}(\vect{r}, \vect{r}) \approx (-)^n 
\frac{2^{n-7/2}z_0^{1/2}}{\pi^{3/2} (Bx^2)^{n-5/2} x^2} 
(1-{z_0}^2)\nonumber\\
\times \int_0^{z_0} \rmd z_-
  \frac{({z_0}^2 - {z_-}^2)^{3(n-3/2)} (1-{z_-}^2)^{1/2}}{({z_0}^4{z_-}^2
  + {z_0}^2{z_-}^4 -4{z_0}^2{z_-}^2 + {z_0}^2 + {z_-}^2)^{3(n-1)/2}}
  \nonumber\\
  \times \exp\left[-\frac{Bx^2}{2} \frac{z_0(1-{z_-}^2)}{{z_0}^2 -
  {z_-}^2}\right].
\end{eqnarray}
A final transformation of variables, by setting $z_- = z_0 
\sqrt{y}/\sqrt{1+y}$, leads to the following asymptotic expression for 
$G^{(n)}_{\perp,\beta}$ in the regime of large $\sqrt{B}\,x$:
\begin{eqnarray}
\fl  G^{(n)}_{\perp,\beta} (\vect{r}, \vect{r}) \approx (-)^n \frac{2^{n-9/2}
    {z_0}^{3n-9/2}(1 -{z_0}^2)}{\pi^{3/2} (Bx^2)^{n-5/2} x^2}  
\exp\left[-\frac{Bx^2}{2 z_0}\right] \int_0^\infty \frac{\rmd y}{\sqrt{y(1+y)}}
  \nonumber\\ 
\times \frac{\sqrt{1+(1-{z_0}^2)y}}
{[2y^2(1-z_0^2)^2+y(1-z_0^2)(3-z_0^2)+1]^{3(n-1)/2}}\nonumber\\
\times\exp\left[-\frac{(1-{z_0}^2)y}{2 z_0}Bx^2 \right].
\end{eqnarray}

The expression found here looks very similar to (\ref{eq:G_1}), which is valid 
for $G^{(1)}_{\perp,\beta}$. As before, we may use the fact that only small 
values of $(1-z_0^2)y$ contribute to the integral for large $\sqrt{B}\,x$. As 
a result, one has the asymptotic relation
\begin{equation}
  \label{eq:higher_reflections}
  G^{(n)}_{\perp,\beta} (\vect{r}, \vect{r}) \approx \left(- \frac{2 {z_0}^3}{B
  x^2}\right)^{n-1} G^{(1)}_{\perp,\beta} (\vect{r}, \vect{r})
\end{equation}
for large $Bx^2$. A similar connection formula holds between the asymptotic 
forms of the excess particles density in various orders:
\begin{equation}
\delta \rho^{(n)}(x) \approx \left(-\frac{2 {z_0}^3}{Bx^2}\right)^{n-1}
 \delta \rho^{(1)}(x)
\end{equation}
again for large $Bx^2$. Likewise, one derives for the asymptotic forms of the 
current density in various orders:
\begin{equation}
  j_y^{(n)}(x) \approx \left(-\frac{2 {z_0}^3}{Bx^2}\right)^{n-1} 
j_y^{(1)}(x).
\end{equation}
We may draw the conclusion that for large $\sqrt{B}\,x$ the $n=1$ term in the 
resummed PDX series yields the dominant contribution, both for the excess 
particle density and for the current density.

\end{document}